# X-RAY STUDY OF THE DENSITY WAVE INSTABILITY OF α- (BEDT-TTF)$_2$MHg(SCN)$_4$ with M=K and Rb

**P. Foury-Leylekian, S. Ravy, and J.-P. Pouget**
Laboratoire de physique des solides, CNRS-UMR 8502, bât. 510
Université Paris-sud, 91405 Orsay Cedex, France

**H. Müller**
European Synchrotron Radiation Facility, BP220
38043 Grenoble, France

We present an X-ray diffraction study of the quasi-2D conductors α–(BEDTTTF)$_2$MHg(SCN), with M=K and Rb. They exhibit a phase transition of the density wave type at $T_{DW}$=8-10K and 12-13K respectively, evidenced by magnetoresistivity, specific heat, NMR and Hall constant measurements. The structural study shows the presence of satellite reflections already at ambient temperature. The related modulation is incommensurate with multiple harmonics. For some of the compounds studied, the intensity of the satellite reflections strongly increases below $T_{DW}$. According to Fermi surface (FS) calculations, the wave vector of the structural modulation achieves a quite good nesting of the global FS. This suggests a coupling of the modulation with the electronic degrees of freedom leading to a charge density wave ground state.

## Introduction

Low dimensional organic charge transfer salts form correlated electronic systems where the balance between intrasite (U) and intersite (V) Coulomb repulsions and electron-phonon coupling stabilize the ground states such as Charge and Spin Density Waves (CDW and SDW), Anti-Ferromagnetism (AF), Spin Peierls (SP) pairing or Superconductivity. Recently, an intriguing mixed 2$k_F$ CDW-SDW ground state has been discovered in the quasi-one dimensional (1D) (TMTSF)$_2$X (X=PF$_6$, AsF$_6$) salts [1]. Since then, numerous theoretical [2] and experimental works have been performed to investigate this density-wave phase. In the quasi-2D α–(BEDTTTF)$_2$MHg(SCN)$_4$ (M=K, Rb) salts, transport and magnetic properties suggest the occurrence of a similar 2$k_F$ SDW-CDW ground state. This paper presents X-ray scattering measurements performed in order to precise the role of the structural degrees of freedom in the low temperature electronic transition observed in these salts.

## Properties of the α phases

α-(BEDTTTF)$_2$MHg(SCN)$_4$ (M=K, Rb, Tl, NH$_4$) phases form a series of isostructural quasi-2D conductors. Its structure [3] (of $P\bar{1}$ space group) is constituted of molecular donor BEDT-TTF layers, separated by thick anionic MHg(SCN)$_4$ sheets. The conducting (a,c) layers are made of **c** stacks of tilted BEDT-TTF molecules with a 2D chevron-like arrangement.

Extended Hückel band structure calculations show that the Fermi Surface (FS) is constituted of 1D open sheets perpendicular to the **a** direction and 2D cylindrical pockets running along the **b** axis [4]. Magnetoresistivity measurements [5] estimate the pocket aera at about 16% of the Brillouin Zone.

Transport and magnetic measurements show the occurrence of a density wave-like transition below $T_{DW}$=8-10K and 12-13K for the K and Rb salts ,respectively. The latter is characterised by a smooth anomaly in the specific heat [6], a hump of resistivity particularly enhanced under magnetic field [7] and a steep increase of the Hall constant. A decrease of the ESR spin susceptibility below $T_{DW}$ and an anomaly at $T_{DW}$ in the NMR relaxation rate [8] were also observed. The magnetic character of the ground state is not clear : neither a broadening nor a splitting of the NMR spectra have been detected below $T_{DW}$.

## Experimental results

X-ray diffuse scattering investigations have been performed using the $\lambda_{Cu}$=1.542Å radiation issued from either a classical tube or a rotating anode equipped with a doubly bent graphite monochromator. The investigation was performed with the so-called fixed film-fixed crystal method. The diffraction set-up was equipped with a Helium cryostat operating from 300K down to 6K. The intensity of selected reflections has been measured with an Ar-methane gas linear detector. The samples studied were thin single crystals (1-2 mm$^2$ by 0.1 mm).

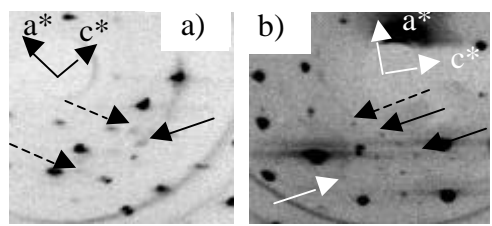

Fig. 1: X-Ray patterns of α (BEDTTTF)$_2$MHg(SCN)$_4$ M=Rb(a) and M=K(b) at 10K. $q_1$ ($q_1$'), $q_2$ ($q_2$') and $q_3$ reflections are shown by plain black, dashed black and plain white arrows respectively.

*a) The potassium salt*

Figure 1b presents an X-ray pattern of the α-(BEDTTTF)$_2$KHg(SCN)$_4$ salt at 10K. The pattern shows in addition to main lattice reflections several weak satellite reflections located at $q_1$=(0.2(1),?,0.45(10)), $q_2$=(0.4(1),?,0.1(1)) and $q_3$=(0.5(1),?,0.5(1)) reduced wave vectors. Accurate measurement performed with a diffractometer have confirmed the $q_1$ wave vector indexation as $q_1$=(0.13(2),0.1(1),0.42(2)). Within the error bars, one has $q_2 \sim 2q_1$ and $q_3 \sim 3q_1$ which means that the $q_2$ and $q_3$ modulations could be the second and third harmonics of a non sinusoidal incommensurate modulation of fundamental wave vector $q_1$. These satellite reflections are already detected at high temperature : up to 250K to 300K ($T_0$) depending on the sample investigated.

The intensity of the satellite reflections, a quantity related to the amplitude of the modulation, has been accurately measured. The intensity is the same for all the $q_1$, $q_2$ and $q_3$ reflections, which shows the strong non sinusoidal character of the modulation. This intensity is rather weak : $5.10^{-3}$ that of a main Bragg reflection. Figure 2 presents the thermal variation of the intensity of a $q_3$ satellite reflection. Upon slowly cooling the sample from room temperature, the intensity first presents a plateau from $T_0$ to 10K and then sharply increases below Ts~10K down to 6K. Ts amounts to about $T_{DW}$. The increase of the intensity at low temperature strongly depends on the thermal treatment of the sample : the sharp increase is not observed after a further increase from 6K to an intermediate temperature $T<T_0$ or when the system is quenched from 300K to 6K. The low temperature satellite intensity increase has not been observed in all the samples investigated.

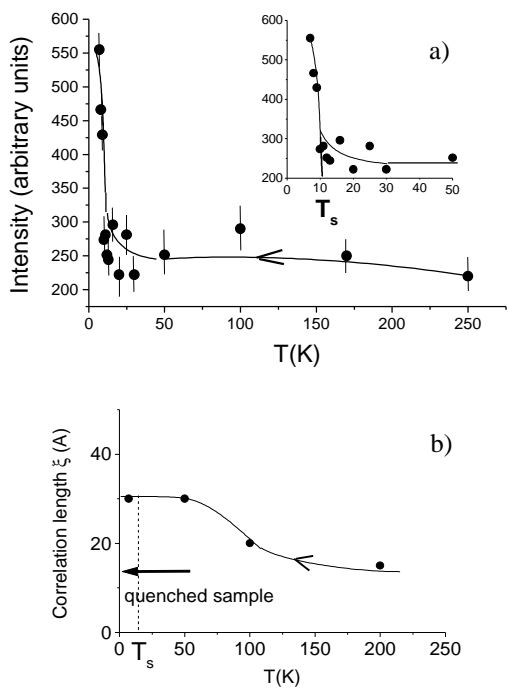

Fig.2 : K salt : a) Thermal variation of intensity of a $q_3$ satellite reflection. The insert represents a zoom at low temperature. b) Thermal variations of the correlation length upon slow cooling. The $\xi$ value obtained after a quench is also indicated.

All satellite reflections widths are generally broader than the experimental resolution. The Lorentzian profile has been observed, characteristic of Orstein-Zernicke type correlations. Figure 3 presents the thermal variations of the in plane correlation length ($\xi$) deduced from the measurement of the half width at half maximum of the $q_3$ satellite reflection ($\xi=\Delta q^{-1}$). The $\xi$ value at 200K corresponds to about 2-3 BEDT-TTF distances. From 200K to 50K, $\xi$ increases and saturates below 50K at a value corresponding to 6 intermolecular distances. Moreover upon quenching the system from 300K to 6K, the satellite reflections remain broad. The $\xi$ value is sample dependant. In one sample investigated, the satellite reflections were found to be quite sharp (fig. 1b).

*b) The rubidium salt*

The X-Ray pattern of the Rb salt taken at 10K (fig1a) shows diffuse spots located roughly at $q_1'=(0.2(1),?,0.3(1))$ and $q_2'=(0.5(1),?,0.5(1))$. These spots are observed in all the temperature range investigated from 300K to 6K. They never condense into satellite reflections. There is always a short range order of the modulation. Within the error bars, the $q_1'$ and $q_2'$ wave vectors are equal to the $q_1$ and $q_3$ wave vectors of the K salt respectively.

**Discussion**

Our investigation shows that the structure of the $\alpha$ (BEDT-TTF)$_2$MHg(SCN)$_4$ (M=Rb and K) salts is modulated even at room temperature. The modulation is triply incommensurate with multiple harmonics, and short range ordered in most of the samples.

In some samples, the amplitude of the structural modulation is enhanced below $T_{DW}$ suggesting a coupling with the electronic degrees of freedom. In that case, we propose that the modulation first affects the anionic group and then extends to the molecular stack below $T_{DW}$, a scenario sustained by the observation of a modification of the g-tensor at $T_{DW}$ [8].

The CDW nature of the $T_{DW}$ transition is evidenced by our investigations. The fact that the **a\*** component of the $q_1$ wave vector well nests the 1D portions of the FS are indeed the fingerprints of a CDW transition. In addition parts of the 2D pockets of the FS are also nested by $q_1$, which achieves the nesting of the global FS [10]. However, the exact mechanism of stabilization of $q_1$ in the high temperature range is still to be clarified.

On the other hand, the SDW character of the transition is questionable. Indeed the only evidence for a magnetic ground state is the change of anisotropy of the total susceptibility at $T_{DW}$. However, this feature is not observed in the spin susceptibility [8]. Moreover, the transition behaves like a CDW under a strong magnetic field [9]. The unconventional features of the transition could be rather related to the absence of a symmetry breaking (the modulation is already present in the high temperature metallic phase) than to the presence of a magnetic component in the density wave modulation.

The difficulty to obtain a long range ordering of the structural modulation, the sample dependence and the kinetics effects are probably related to chemical or structural disorder. It is known that a CDW is very sensitive to such a disorder. In the present case, it can easily destroy a loose nesting of the FS. Such features could explain the difference of physical properties between nearly isostructural compounds. In this respect, for the NH$_4$ salt for which a superconductor ground state is stabilized, preliminary investigations do not reveal the presence of a structural modulation as for the K and Rb salts.